\documentclass{PoS}
\usepackage{amsmath}
\usepackage{url}
\usepackage{graphicx}
\title{CRIME - cosmic ray interactions in molecular environments}

\ShortTitle{CRIME}

\author{\speaker{Julian Krause}\\
        APC Universite Paris Diderot, 75014, Paris, France\\
        E-mail: \email{julkrau@googlemail.com}}
\author{Giovanni Morlino\\
        Gran Sasso Science Institute, viale F. Crispi 7, I-67100 L’Aquila, Italy \\
        E-mail: \email{giovanni.morlino@gssi.infn.it}}        
\author{Stefano Gabici\\
        APC Universite Paris Diderot, 75014, Paris, France\\
        E-mail: \email{gabici@apc.in2p3.fr}}

\abstract{
Molecular clouds act as targets for cosmic rays (CR), revealing their presence through either gamma-ray emission due to proton-proton interactions, and/or through the ionization level in the cloud, produced by the CR flux.
The ionization rate is a unique tool, to some extent complementary to the gamma-ray emission, in that it allows to constrain the CR spectrum especially for energies below the pion production rate ($\approx 280$ MeV).
Here we study the effect of ionization on $H_2$ clouds due to both CR protons and electrons, using the fully relativistic ionization cross sections, which is important to correctly account for the contribution due to relativistic CRs. The contribution to ionization due to secondary electrons is also included self-consistently.
The whole calculation has been implemented into a numerical code which is publicly accessible through a web-interface. The code also include the calculation of gamma-ray emission once the CR spectrum (electrons and/or protons) have been specified.}

\FullConference{The 34th International Cosmic Ray Conference,\\
		30 July- 6 August, 2015\\
		The Hague, The Netherlands}

\begin{document}

\section{Introduction}
In this work we present a tool to model the ionization caused by cosmic rays (CRs) as a complementary tracer to gamma ray observations.
CRs are deflected by interstellar magnetic fields due to their charge and thus loose all the information about their origin. To observe cosmic rays in the vicinity of their sources one therefore relies on 
tracers of cosmic rays. The most commonly known are gamma rays produced in cosmic ray interactions with matter or radiation fields. Since the most dominant species of CRs are protons the sources of cosmic ray protons are the dominant sources of cosmic rays. Protons prodcue gamma-rays in inelastic proton proton interactions via the production and subsequent decay of 
neutral pions. Therefore, molecular clouds in the vicinity of cosmic ray sources act as barometers for the local cosmic ray flux (dependning on the mass and distance to the molecular cloud). In fact observations in high and 
very high energy gamma rays revealed proton acceleration in supernova remnants interacting with molecular clouds \cite{fermi, agile}. 
In this work we investigate a CR tracer in molecular clouds complementary to gamma rays. Here we investigate the ionisation rate of CR potons and electrons. In contrast to the gamma ray emission, the ionisation 
rate is a single quantity integrated over the whole CR spectrum. Therefore, the ionisation rate does not contain direct information on the underlying CR proton spectrum. 
However, in contrast to gamma-ray emission, the ionization rate level is not limited to proton energies above 280 MeV being produced mainly by lower energies CRs. Therefore cosmic ray ionisation measurements allow to access the low energy end of the cosmic ray proton 
spectrum in interstellar space, offering a unique possibility to measure energies below the pion production threshold. 

Here we present the basic principles that determine the ionisation rate caused by cosmic rays. 
With respect to previous works on this subject, we improved the calculations in two aspects: i) we use the relativistic differential cross section and ii) we include a self-consistent calculation of ionization due to secondary electrons. The use of differetial 
cross sections allows a selfconsistent calculation of the secondary ionizations. 
The calculations presented here have been implemented in a code which is available online (http://crime.in2p3.fr/) and free available for the scientific community. 
The online tool calculates the ionisation rate for a given cosmic ray proton and/or electron spectrum and provides some basic informations such as the most ionizing energies of the considered CR spectrum. 

\section{Cross sections, secondary ionisations and propagation}
\begin{center}
 \begin{figure}
\hspace{-1cm}
\includegraphics[width=0.45\textwidth]{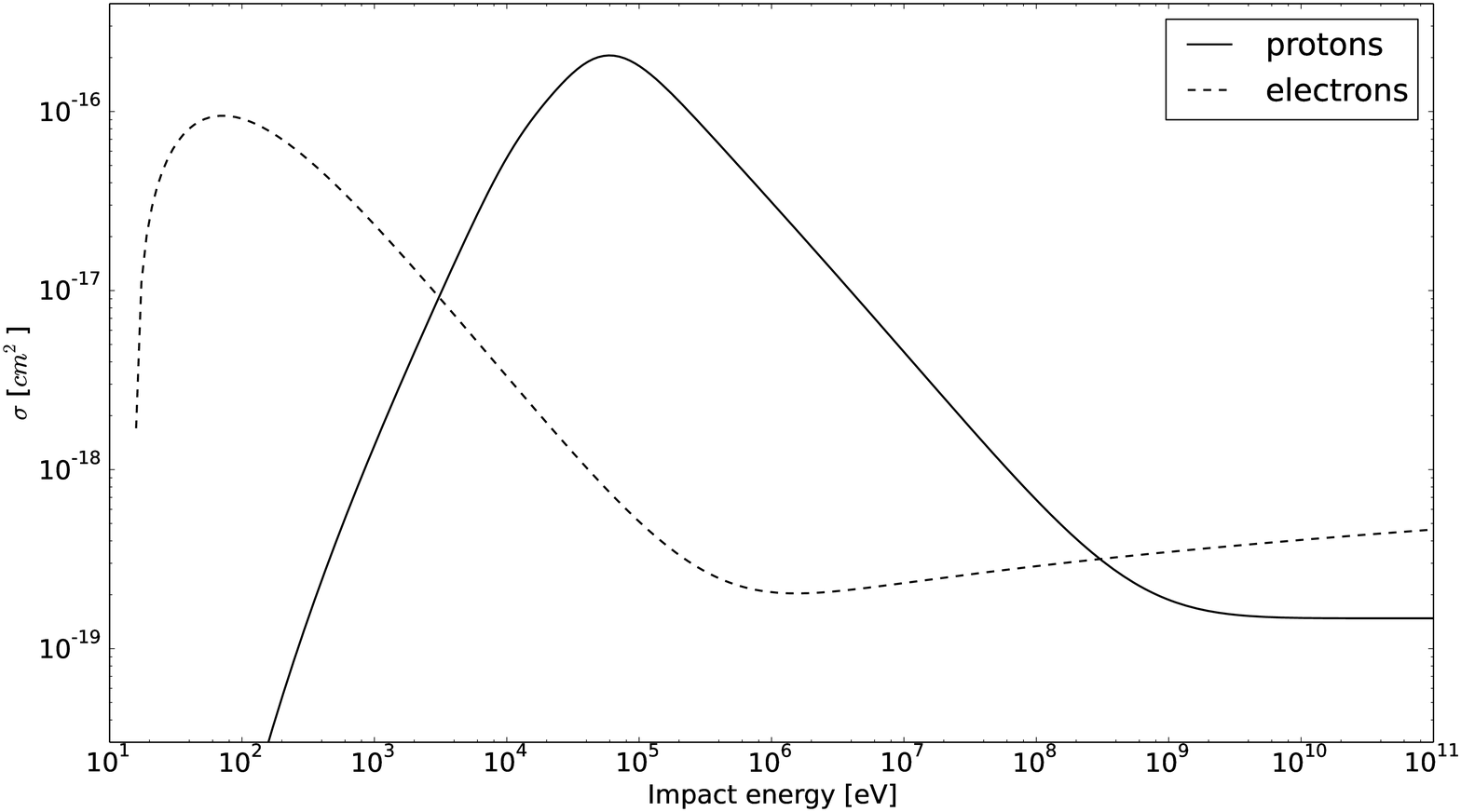}
\includegraphics[width=0.45\textwidth]{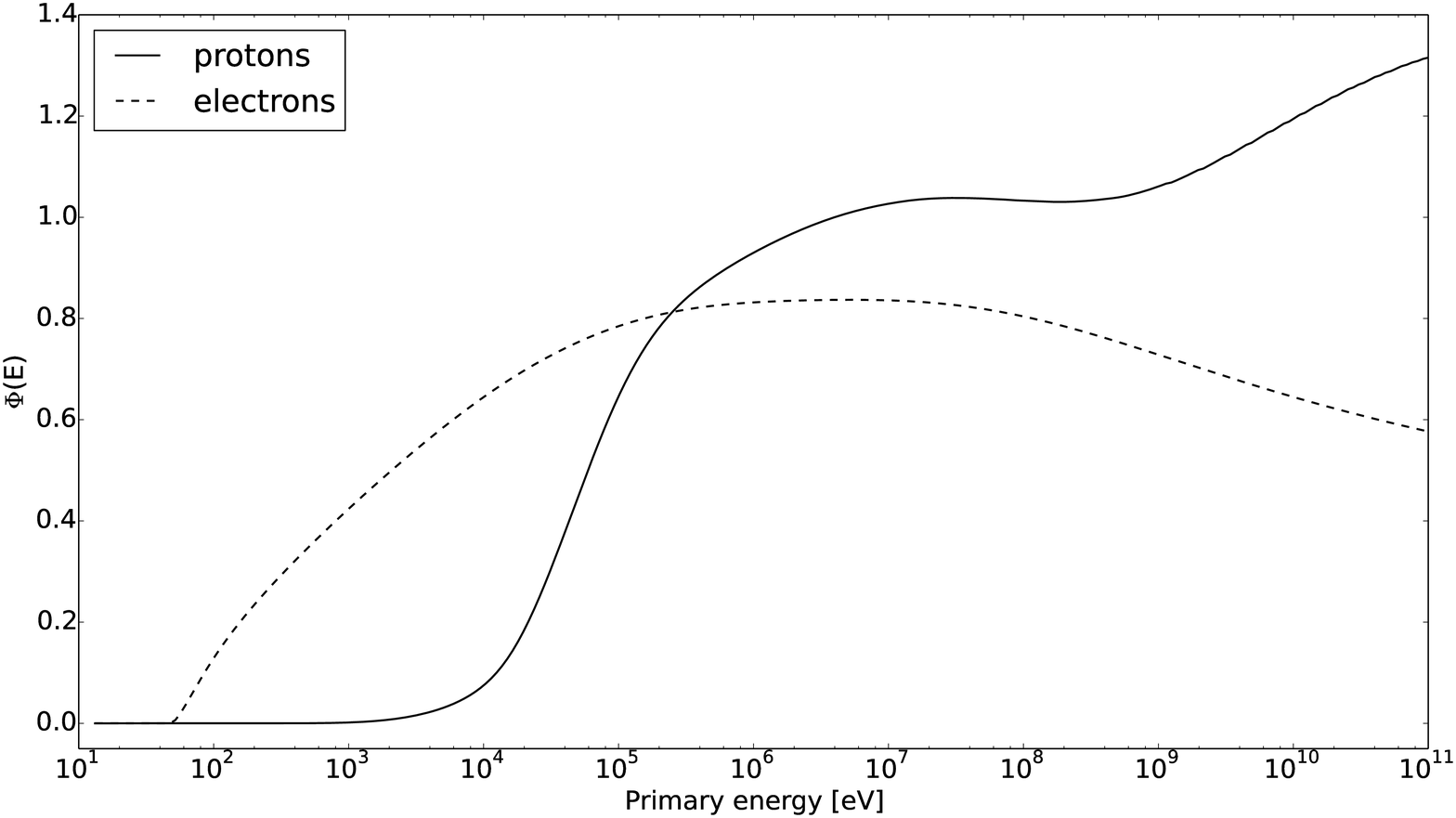}\\
\caption{Shown are the total ionisation cross section (left hand side) and the secondary ionisations per primary ionisation (right hand side) as a function of the energy for both electrons and protons on a pure 
$H_2$ target.}
\label{fig1}
\end{figure}
\end{center}

In the following section we briefly summarize the ionisation cross sections used in our code CRIME (Cosmic Ray Interactions In Molecular Environments). 
In addition we derive a general expression describing the secondary ionisations. Figure~\ref{fig1} shows the cross section 
and secondary ionisations for molecular hydrogen. Finally we shortly adress the issue of cosmic ray propagation into a molecular cloud. These section describes the building blocks of the CRIME online tool. 

\subsection{Electron cross section}
We use the relativistic binary-encounter-dipole model for the differential ionization cross section of electrons. The relativistic differental cross 
section is than given (eq. 19 \cite{kim2000}) as:

\begin{align}
\label{kim_diff}
 \left( \frac{d\sigma}{dW} \right) = &\frac{4\pi a_0^2 \alpha^4 N}{(\beta_t^2 + \beta_u^2 + \beta_b^2)2b'} \nonumber\\
                                     & \left\{ \frac{N_i/N-2}{t+1} \left( \frac{1}{w+1} + \frac{1}{t-w} \right) \frac{1+2t'}{(1+t'/2)^2} \right.  \nonumber\\
                                     &+ \left. [2-(N_i/N)] \left[ \frac{1}{(w+1)^2} + \frac{1}{(t-w)^2} + \frac{b'^2}{(1+t'/2)^2}\right] \right.  \nonumber\\
                                     &+ \left. \frac{1}{N(w+1)} \frac{df}{dw} \left[ ln \left( \frac{\beta_t^2}{1-\beta_t^2} \right) -\beta_t^2 - ln(2b') \right] \right\}
\end{align}

Here $w=W/I$ and $t=T/I$ are the energies in units of the ionization potential 
where $T$ is the energy of the primary electron and $W$ the energy of the secondary electron.
$t_p=T/m$, $b_p=B/m$ and $u_p=U/m$ are the primary energy, binding energy (ionization potential) and average orbital kinetic 
energy of an electron in natural units. 

\begin{align}
 t'=T/mc^2 \,\,\, , \beta_t^2 = 1- \frac{1}{(1+t')^2} \\
 b'=B/mc^2 \,\,\, , \beta_b^2 = 1- \frac{1}{(1+b')^2}\\
 u'=U/mc^2 \,\,\, , \beta_u^2 = 1- \frac{1}{(1+u')^2}
\end{align}

The dipole oscillator strength $\frac{df}{dw}$ depends on the target material and can be approximated as a power law series for the materials under consideration.

\begin{equation}
 \frac{df}{dw}= \frac{c}{(1+w)^3} + \frac{d}{(1+w)^4} + \frac{e}{(1+w)^5} + \frac{f}{(1+w)^6}
\end{equation}

For $H_2$ the fitting parameters are:  c=1.1262,  d=6.382,  e=-7.8055, and  f=2.144. 

Integrating equation \ref{kim_diff} over the secondary electron energy ($w$) yields the total ionization cross section (see also eq. 20 in \cite{kim2000}):
While the integral is partially analytical, the part of the integral involving the dipole oszilator strength $\frac{df}{dw}$ can be isolated and is given as: 

\begin{equation}
 D(t) \equiv N^{-1} \int_0^{(t-1)/2} \frac{1}{w+1} \frac{df}{dw} dw.
\end{equation}

For the cases considered here, where the dipole oszilator strength can be described as a simple powerlaw series, the above integral 
can also be solved analytically.

\subsection{Proton cross section}
For the proton ionization cross section we adopt an analytical expression given by \cite{Rudd1988}. This expression is based on the molecular promotion model in the low energy regime and on the classical binary encounter approximation for higher energies.

The model equation for the differential cross section as a function of the ejected electron energie ($w=E_e/I$) in units of the ionization potential 
($I$) and the reduced proton velocity ($v=\sqrt{\frac{m_e}{m_p} \frac{E_p}{I}}$) is given as: 

\begin{align}
\label{rudd}
 \frac{d\sigma}{dw}= (S/I) \, \frac{[F_1(v)+F_2(v)w](1+w)^{-3}} {1+\exp[\alpha'(w-w_c)/v]}
\end{align}

with 
\begin{equation}
  S=4\pi a_0^2 N (R/I)^2, 
\end{equation}
\begin{equation}
\label{wc}
 w_c=4v^2-2v-R/4I.
\end{equation}

The parameter $w_c$ determines the point where the exponential surpression of the cross section sets in and is in fact composed of two parts, 
$w_c=A-B$ with $A=4v^2-2v$ and $B=R/4I$. As long as $B\gg A$ we are in the low energy part of the cross section which is described by the molecular 
promotion model. Within the low energy regime $w_c$ is independent of the energy of the impacting proton and determined by the target material.

For impact energies $A \gg B$ the cross section is based on the binary encounter model. The maximum of the total cross section lies 
in the range where $A\sim B$ ($E\sim 10^5$). The here used definitions of $A$ and $B$ are the ones used in \cite{Rudd1988}. It is worth to mention that 
a more general definition of

\begin{equation}
\label{A}
 A=4E-\gamma \sqrt{I E}
\end{equation}

with $\gamma=4$ is given in the binary encounter theory. However, \cite{Rudd1988} report that $\gamma =2$ has been found experimentally and is thus the value used in 
the model equation~\ref{rudd}. In the classical regime ($E=1/2 mv^2$) and with energy measured in units of the ionisation potential $I$, one obtains the 
expression used in equation~\ref{wc}. $F_1$ and $F_2$ are described by the following fitting functions:
\begin{equation}
 F_1=L_1+H_1 \,\,\,  \rm{and}  \,\,\, F_2=L_2 H_2 /(L_2+H_2)   
\end{equation}
\begin{align}
H_1 &= A_1 \ln(1+v^2)/(v^2+B_1/v^2), \,\,\,  \rm{and}  \,\,\,  H_2 = A_2/v^2 + B_2/v^4, \\
L1  &= C_1 v^{D_1} / (1+E_1 v^{D_1+4}), \,\,\,  \rm{and}  \,\,\, L2  = C_2 v^{D_2}
\end{align}. 

The ten parameters $A_1,...,E_1$, $A_2,...D_2$ and $\alpha'$ are obtaind by fitting the model equation to the data. In the case of $H_2$ as a target they are: $A_1=0.80$,  $B_1=2.9$, 
$C_1=0.86$, $D_1=1.48$,  $E_1=7.0$, $A_2=1.06$, $B_2=4.2$, $C_2=1.39$, $D_2=0.48$, and $\alpha'=0.87$ \cite{Rudd1988}. To obtain the total cross section equation~\ref{rudd} is numerically integrated.

\paragraph{Relativistic corrections to the Rudd model}
The model equation of \cite{Rudd1988} is not correct for relativistic proton energies. To expand the model to relativistic impact energies we change the 
velocity parameter $v$ to hold in the relativistic limit. First we compare the reduced velocity parameter with the classical velocity $v_cl$: 

\begin{equation}
 v_{cl}=\sqrt{2E/m_p}.
\end{equation}

We find the relation
\begin{equation}
   v=\sqrt{m_p/2} \times \sqrt{m_e/(m_p/I)} \times v_{cl}.  
\end{equation}

However, in the proper relativistic treatment the velcovity is given as: 

\begin{equation}
 v_{rel}=\sqrt{1.-(m_p c^2/(E+m_p c^2)^2} \times c
\end{equation}

where $c$ is the speed of light. We therefore substitute the velocity parameter $v$ with

\begin{equation}
 v'=\sqrt{m_p/2} \times \sqrt{m_e/(m_p/I)} \times v_{rel}
\end{equation}

to introduce a velocity parameter which holds in the relativistic limit and reduces to the old parameter for not relativistic energies. 
Moreover, we can not use Equation~\ref{wc} as it was derived using the classical energy-velocity relation. Analog to the above derivation we find that:

\begin{equation}
 w_c'=(4E-2\sqrt{IT})/I-w2.
\end{equation}

These changes give a sufficiently correct approximation of the cross section in the relativistic limit. Due to the above changes the total ionization cross section stays constant for highly relativistic energies as the normalization of 
the differential cross section depends only on the velocity which asymptotically approaches $c$. In reality the cross section keeps increasing logarithmically in energy, introducing a small uncertainty for highly 
relativistic particles.  

\subsection{Secondary ionisations}
Each ionization process produces a free electron(here we neglect the case where both electrons are stripped from the $H_2$ molecule). Those secondary electrons can produce further ionizations if their kinetic energy is above the ionisation threshold. 
In this paraghraph we want to determine the effect of ionisation from secondary electrons as a function of the primary ionization process. Given 
the differential cross section the production rate of secondary electrons is: 

\begin{equation}
 \frac{dN_s}{dEdVdt}= 4\pi \int_I^{\infty} n \, j'(E') \frac{d\sigma'(E,E')}{dE} dE' 
\end{equation}
Here $n$ is the number density of target molecules, $j'$ is the flux of primary particles and $\frac{d\sigma'(E,E')}{dE}$ is the differential 
ionization cross section for the primary particles. In the following \textit{prime} parameters always refer to the quantities of the primary particles.
Assuming a steady state solution the spectrum of secondary electrons is given as:

 \begin{equation}
  \frac{dN_s}{dEdV}= \frac{dN_s}{dEdVdt} \times \tau 
 \end{equation}

Where $\tau$ is the residence time of secondary particles within the region of interest. We shall assume that all secondaries do not travel any 
significant distance from the point of origin before loosing all their energy. In this case the residence time corresponds to the energy-loss time, 
and all secondaries exist only at the location where the primary ionization took place. Rewriting the energy loss function as $L(E)$, the energy 
loss per column density, the residence time is given as:

\begin{equation}
 \tau = \tau_{loss} = \frac{E}{dE/dt}=\frac{E}{L(E)nv}  
\end{equation}
Where $n$ is the number density of the target material responsible for the energy losses and $v$ the velocity of the secondary electrons. For simplicity we 
consider a molecular cloud consisting of molecular hydrogen only. The energy loss function is taken from \cite{Padovani2009} based on measurements. 
Therefore it accounts for all relevant processes, namely: ionisation losses, 
losses due the excitation electronic levels and vibrational states, as well as Bremsstrahlung. Using the above expression for the energy losses and the spectrum of 
secondaries we obtain the flux of secondary electrons assuming isotropy as: 

\begin{align}
  j_s(E) &= \frac{dN_s}{dAdEdtd\Omega} = \frac{v}{4\pi} \frac{dN_s}{dEdV}   \nonumber \\[5pt]
         &=\frac{E}{L(E)} \int_I^{\infty}  j'(E') \frac{d\sigma'(E,E')}{dE} dE' 
 \end{align}

The flux of secondary electrons is proportional to the ratio of production rate to loss rate as expected in a steady state solution. Now we can calculate the ionization 
rate for the secondaries as we did for the primaries. 
To better understand the effect of secondaries, it is convenient to define a function, $\Phi(E')$, as the average number of secondary ionizations per ionization event of a primary with energy $E'$. 

\begin{align}
	 \Phi(E') &\equiv \frac{\int_I^{E_{max}} \frac{j_s(E)}{dE'} \sigma_s(E)dE} {j'(E')  \, \sigma'(E') }  \nonumber \\[5pt]
		  &= \frac{1}{\sigma'(E')} \int_I^{E_{max}} \frac{d\sigma'(E',E)}{dE} \frac{E \,\sigma_s(E)}{L(E)} \,  dE  
\end{align}

We find that $\Phi(E')$ does not depend on the flux of primaries but only on the total and differential ionization cross section of the 
primary species, and the total ionization cross section and the energy losses of the secondaries (electrons). 
Therefore, for a given target material and a given energy of primary particle, $\Phi(E)$ is a constant.
We note that for multi component targets the definition of $\Phi(E')$ has to be 
understood as a function of multiple components where $\Phi(E')_{t_p, t_s}$ is the number of secondary ionizations of target species $t_s$ caused by 
primary ionizations of target species $t_p$. 
Interstellar molecular clouds are heavily dominated by molecular hydrogen which shows a higher ionisation cross section than other abundant targets (like Helium). 
Therefore, secondaries produced on any other target than molecular hydrogen will be less in number and mainly lead to further ionization 
of molecular hydrogen. For the same reasons also the energy loss of secondary electrons is dominated by molecular hydrogen. 
As can be seen from the right panel of Figure 1, for H2 $\Phi(E')$
 is of the order of unity for both electrons and protons and thus represents an important correction. 

\subsection{Propagation of cosmic rays into a molecular cloud}
To estimate the ionisation deep inside a molecular cloud it is important to correctly describe the propagation of CRs inside the cloud and to account for their energy losses. While this in general a complex field of research we follow 
\cite{Padovani2009} and assume straight line propagation and continuous energy losses along the line of sight. This reflects the minimum amount of energy losses to any CR spectrum. It is important to note that even though 
CRs below a certain energy $E_s$ can not penetrate beyond a certain column density, a low energy tail of the CR spectrum below $E_s$ is present at any depth. This low energy tail is produced by CRs of energies  
$E_s +\epsilon$ and its shape is given by the inverse energy loss function (see \cite{Padovani2009}) for details). The importance of the low energy tail for the total ionisation rate depends on the shape of the 
assumed CR spectrum outside the cloud and the penetration depth considered.

\section{CRIME the online tool}
\begin{center}
\begin{figure}
\includegraphics[width=0.45\textwidth]{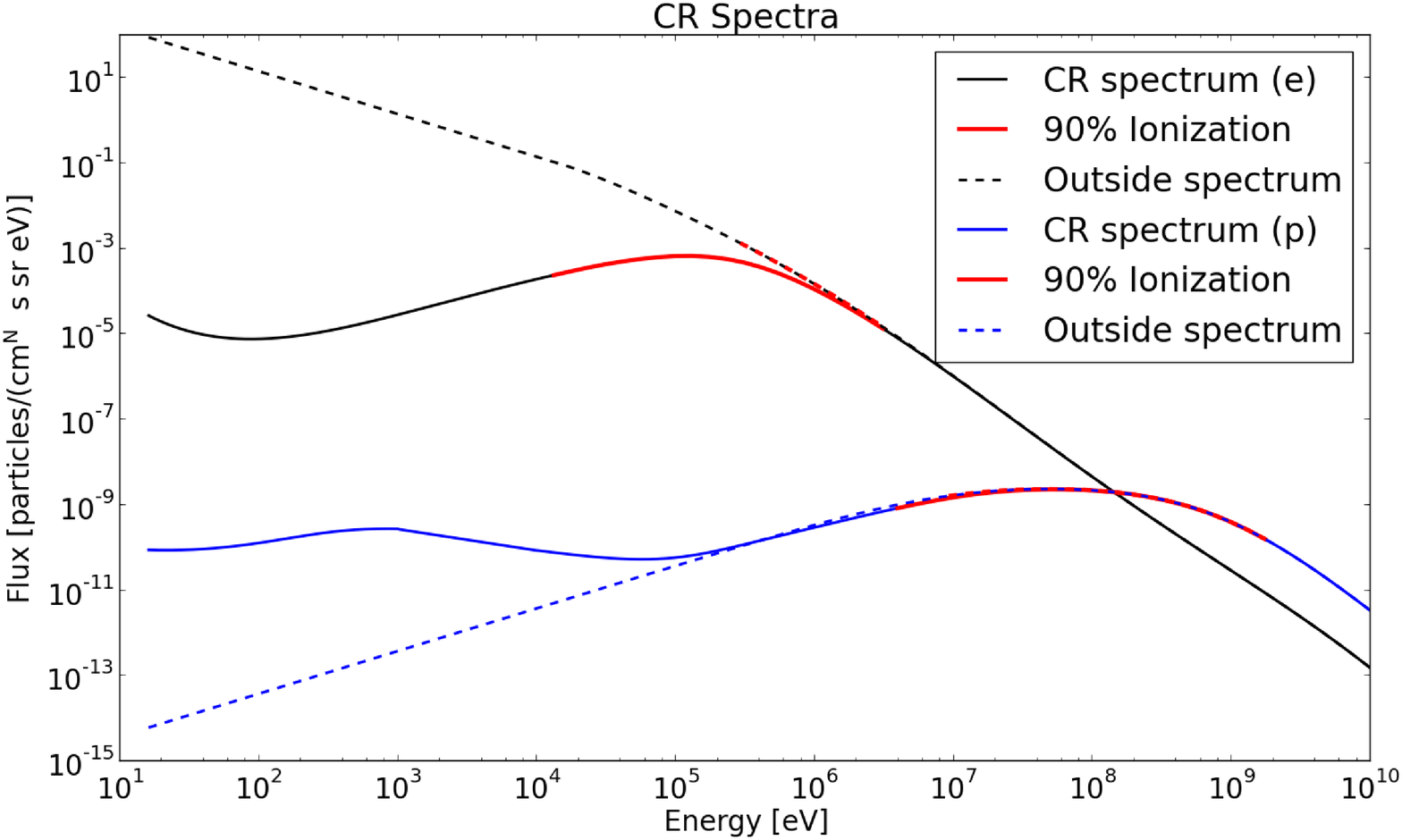}
\includegraphics[width=0.45\textwidth]{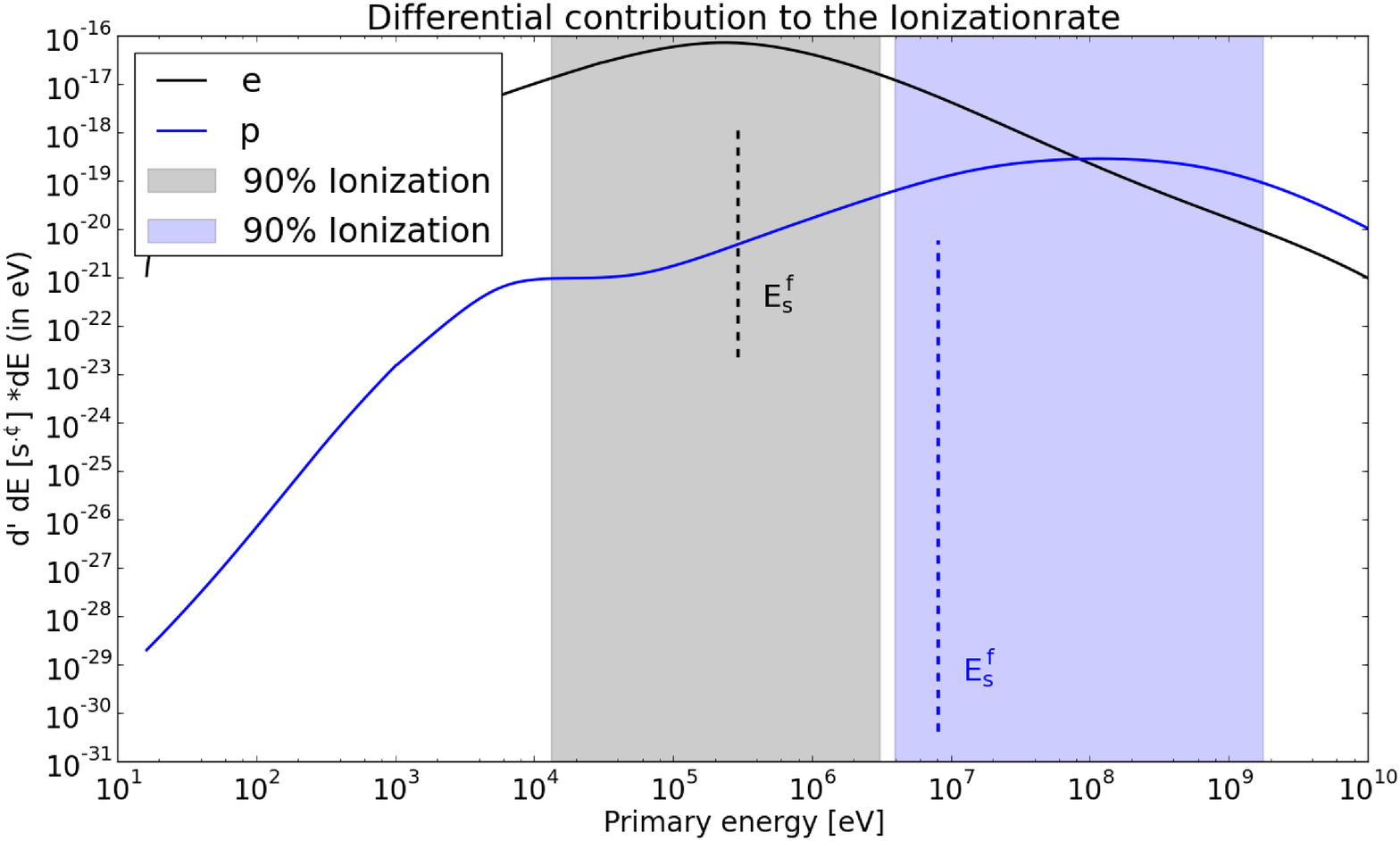}\\
\caption{Example Output of a CRIME run. Left hand side: Shown are the CR input spectra for electrons and protons before and after straight line propagation to a column density of $H_2$ of $10^{22} \rm{cm}^2$. 
The part of the spectra marked in red contribute to 90\% of the ionisation rate. The right hand side shows the differential contribution to the ionisation rate in arbitrary units. The regions contributing to 90\% of the 
ionisation are highlighted and in addition the energy $E_s$ not able to penetrate to the chosen $H_2$ column density of $10^{22} \rm{cm}^2$ is shown. We note that for electrons the low energy tail of the spectrum 
produced in the cooling process accounts for about 50\% of the ionisation, while for protons its contribution is minimal.}
\label{fig2}
\end{figure}
\end{center}

CRIME is a webrun tool accessible under \url{http://crime.in2p3.fr/webrun.html} that allows the calculation of the CR ionistion rate of $H_2$. 
The user can specify an electron and/or proton spectrum among a power law or broken power law shape as well as pre-defined minumim and maximum CR background models. In the near future the choice of arbitrary CR spectra 
will be implemented via a file upload (data points). It is also possible to account for the straight line propagation of the spectrum to the column density of interest, alternatively a minimum CR energy considered 
can be specified. Based on the user choices, the calculation and the displaying of the results takes 10-40 seconds. The output of the calculation is the ionisation rate per second as well as two plots showing the user 
defined CR spectrum (and the propagated spectrum if requested) and the differential contribution to the ionisation rate. The region contributing to 90\% of the total ionisation rate is highlighted. In case the 
straight line propagation is calculated, also the shielded energy $E_s$ is calculated and marked in the plot showing the differential contribution to the ionisation rate. An example output for a \textit{Voyager}-like 
CR proton model and a maximum CR background electron model is shown in Figure~\ref{fig2}. In this example the ionisation rate per second is $1.8\times10^{-17}$ and $3.6\times10^{-15}$ for protons and electrons respectively.  
This tool will provide a fast and easy access to the calculation of the CR ionisation rate as well as basic information like the most relevant energies causing it. 

\vspace{-1cm}

\section{Conclusion and Outlook}
We provide a easy-to-use web tool to calculate the ionisation rate of CR electrons and protons based on relativistic cross sections and a selfconsistent calculation of the effect of secondary ionisations.
Straight line propagation of the CR spectra is available as a minimum energy loss calculation. In addition to the value of the ionisation rate the tool provides information on the effect of the propagation model as well
as the differential contribution to the ionisation rate. 

We are currently working on the implementation of the calculation of the gamma-ray emission of the specified cosmic ray spectrum (with the pion production channel already available on the webpage) to allow a 
simultaneous calculation of ionisation rate and gamma-ray emission. The whole calculation is based on a very modular code which allows an easy implementation of new features like more complicated propagation models or 
other target materials for the ionisation (with Helium being already implemented in the underlying code).


\begin{thebibliography}{99}
\footnotesize
\bibitem{kim2000}
Y.-K. {Kim}, J.~P. {Santos}, and F.~{Parente}.
\emph{Extension of the binary-encounter-dipole model to relativistic incident electrons}.
\emph{Physical Review A} (Atomic, Molecular, and Optical Physics), Volume 62, Issue 5, 2000

\bibitem{Padovani2009}
Padovani, M.; Galli, D.; Glassgold, A. E.
\emph{Cosmic-ray ionization of molecular clouds}
\emph{Astronomy and Astrophysics, Volume 501, Issue 2, 2009}

\bibitem{Rudd1988}
Rudd, M. E.
\emph{Differential cross sections for secondary electron production by proton impact}
\emph{Physical Review A (General Physics), Volume 38, Issue 12, December 15, 1988}

\bibitem{fermi}
Ackermann, M. et al.
\emph{Detection of the Characteristic Pion-Decay Signature in Supernova Remnants}
\emph{Science, Volume 339, Issue 6121, 2013}

\bibitem{agile}
Giuliani, A. et al.
\emph{Neutral Pion Emission from Accelerated Protons in the Supernova Remnant W44}
\emph{The Astrophysical Journal Letters, Volume 742, Issue 2, 2011}
\end{thebibliography}
\end{document}